\documentclass[
 preprint,
 graphics,
 floatfix,
 tightenlines,
 superscriptaddress,
 nofootinbib,
 nobibnotes,
 amsmath,
 amssymb,
 aps,
 prl
]{revtex4-1}

\usepackage[caption=false]{subfig}
\usepackage{amssymb}
\usepackage{amsmath}
\usepackage{commath}
\usepackage{graphicx,bm}
\usepackage{verbatim}

\usepackage{dcolumn}
\DeclareMathOperator{\Tr}{tr}
\usepackage{color}
\newcommand{\ad}[1]{\textcolor{blue}{{\it#1}}}

\usepackage{ulem}

\begin{document}

\title{Two-dimensional, blue phase tactoids}

\author{Luuk Metselaar}
\author{Amin Doostmohammadi}
\author{Julia M. Yeomans}
\affiliation{The Rudolf Peierls Centre for Theoretical Physics, Clarendon Laboratory, Parks Road, Oxford OX1 3PU, UK}

\begin{abstract}
We use full nematohydrodynamic simulations to study the statics and dynamics of monolayers of cholesteric liquid crystals. Using chirality and temperature as control parameters we show that we can recover the two-dimensional blue phases recently observed in chiral nematics, where hexagonal lattices of half-skyrmion topological excitations are interleaved by lattices of trefoil topological defects. Furthermore, we characterise the transient dynamics during the quench from isotropic to blue phase. We then proceed by confining cholesteric stripes and blue phases within finite-sized tactoids and show that it is possible to access a wealth of reconfigurable droplet shapes including disk-like, elongated, and star-shaped morphologies. Our results demonstrate a potential for constructing controllable, stable structures of liquid crystals by constraining 2D blue phases and varying the chirality, surface tension and elastic constants.
\end{abstract}

\maketitle

\section{Introduction}

Recent work has demonstrated the possibility of stabilising exotic topological field configurations, which cannot be deformed into a uniform field by a smooth or continuous process, in diverse physical systems~\cite{Ravnik2014,Ravnik2015,Musevic2017}. These include skyrmions \cite{Skyrme1962}, which comprise 180$^{\circ}$ double twist of an order parameter radially outwards from a centre, half-skyrmions (or merons) \cite{Braun2012}, where the twist is 90$^{\circ}$, and hopfions where the variation in the order parameter is three-dimensional. Such structures are found in chiral magnets, where they have potential applications in spintronic and magnetic memory devices,  in Bose-Einstein condensates due to Rashba spin-orbit coupling \cite{Xu2012} and in antiferromagnets, where the Dzyaloshinsky-Moriya interaction leads to weak ferromagnetism \cite{Dzyaloshinsky1958}. Examples of topological phases formed by ordered arrays of half-skyrmions are the blue phases of chiral liquid crystals \cite{Wright1989,Gleeson2015}. These consist of regular arrays of double twist cylinders, where there is chiral ordering in two directions around a central axis. The double twist cylinders cannot fill space without introducing singularities in the director field, and hence they are interleaved by disclination lines. The blue phases therefore occur close to the isotropic-cholesteric transition, where the energetic advantage of the double twist ordering is sufficient to outweigh the energetic penalty of the defect cores.

\begin{figure}[b]
  \includegraphics[width=1\textwidth]{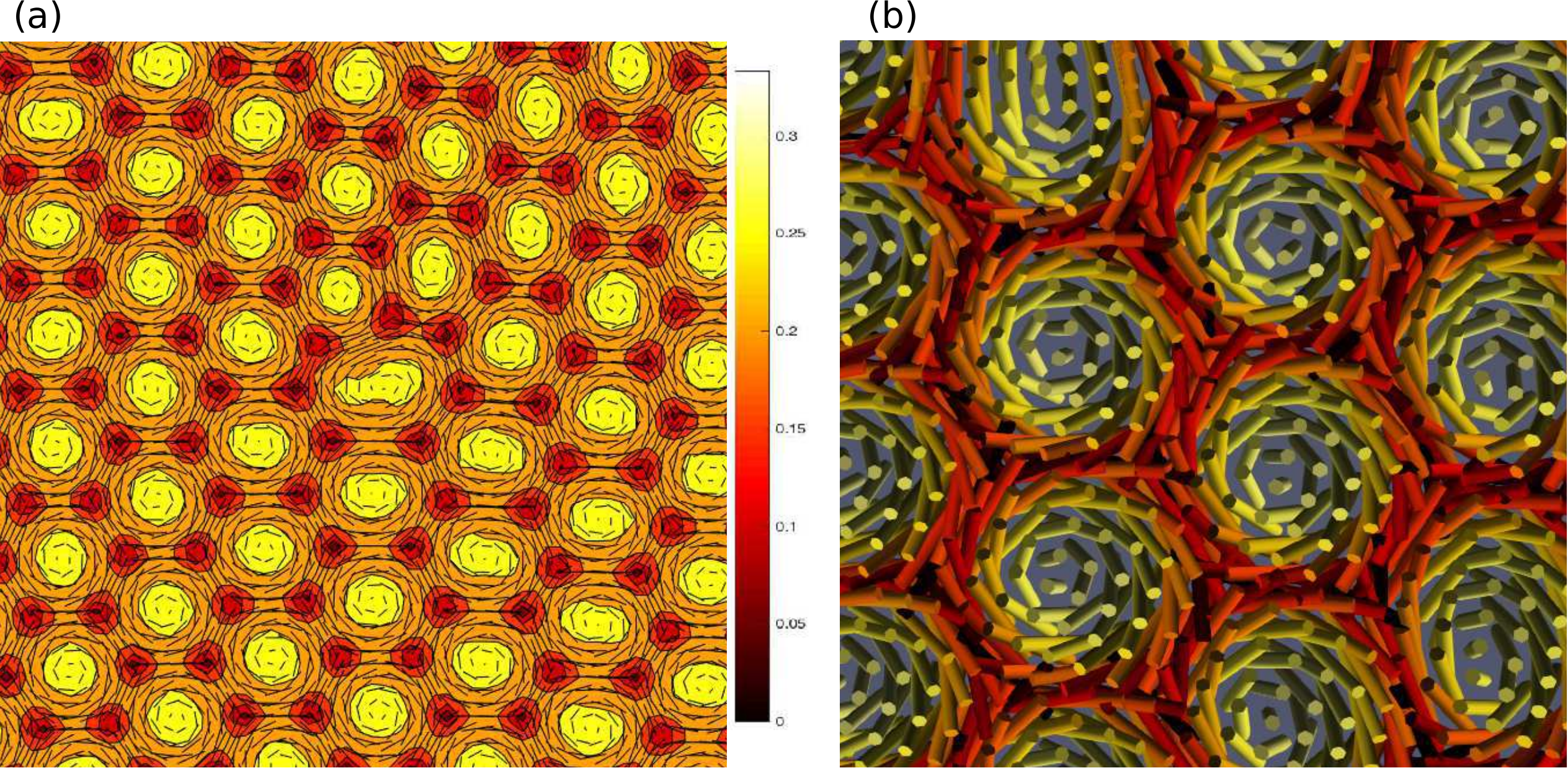}%
\caption{A two-dimensional blue phase with periodic boundary conditions. (a) The colour indicates the magnitude of the nematic order $S$; the half-skyrmions are yellow, and the topological defect cores appear dark red. Black lines are the projections of the director onto the plane. Imperfections in the hexatic order are due to kinetic trapping and finite size effects. (b) Magnified image of the director field. The colour coding indicates the strength of the nematic order, with yellow corresponding to a stronger orientational order.}
\label{fig:example_2dbluephase}
\end{figure}

The possibility of a blue phase in two dimensions was predicted numerically \cite{Fukuda2011}, and has very recently been confirmed experimentally, in liquid crystals confined between two plates with planar degenerate anchoring \cite{Nych2017}. The structure of the 2D blue phase is a hexagonal lattice of double twist regions, (i.e. half-skyrmions), separated by an interpenetrating lattice of $-1/2$ topological defects as shown in Fig.~\ref{fig:example_2dbluephase}. These experiments suggest new possibilities for studying the interplay between the properties and conformations of half-skyrmions and liquid crystalline features and structures. As a step in this direction, here we numerically investigate the confinement of cholesteric and blue phases within lyotropic liquid crystal droplets.

Many lyotropic liquid crystals form nematic droplets, or tactoids, liquid crystal microdomains within the isotropic phase, often as a consequence of phase ordering. Nematic tactoids take a characteristic shape, spherical or spindle-like depending on their size, because of the interplay between elastic and surface tension free energies and interfacial anchoring \cite{Jamali2015,Prinsen03,Metselaar2017}. This suggests the possibility that novel tactoid shapes might form in a system where the bulk ordering is a 2D blue phase. Therefore in this paper we present numerical results demonstrating the structure of 2D chiral nematic tactoids as the pitch, droplet size, elasticity and the surface tension vary.
 
Moving to a longer length scale, colloidal liquid crystals such as finite rafts of fd-virus, have been shown to form interesting structures when on the surface of a liquid. Here the depletion interaction means that the liquid crystal colloids prefer to point perpendicular to the surface and hence half-skyrmions are unfavourable, and the blue phase is replaced by a twist localised around the outside of the droplet \cite{Gibaud2012}. We show that within our model similar energetic effects can be mimicked by an external electric field, and we study the crossover from blue phase to raft-like behaviour.

The paper is organised as follows: first we describe our numerical approach and present a phase diagram to illustrate the region of stability of the 2D blue phase, both in the absence and the presence of an external field. We also describe the dynamics of quenches from the isotropic phase into the blue phase. We next consider 2D chiral tactoids and present results for their shapes and director field configurations. In the limit of a strong field we show that our results reproduce the structures seen in circular membranes of fd-virus.

\section{Simulation method}
We employ a continuum approach based on the nematohydrodynamic equations of chiral liquid crystals to study the morphology and dynamics of chiral tactoids. Within a Landau-de Gennes theory the orientational order of a liquid crystal is described by a second-rank symmetric traceless tensor $\bm{Q}=\frac{3}{2}\mathcal{S}\left(\bm{nn}-\frac{1}{3}\bm{I}\right)$, where $\mathcal{S}$ is the magnitude of the orientational order and ${\bf n}$ is the nematic director. To distinguish between the isotropic and cholesteric phases we use an additional order parameter $\varphi$, which is the liquid crystal concentration. The two phases can coexist, and the amount of each is conserved. The free energy density of the system is 
\begin{equation}
\begin{split}
f = \frac{1}{2}A_{\varphi}\varphi^2\left(1-\varphi^2\right) + A_0\varphi^2\Big\{\frac{1}{2}\Big(1-\frac{\eta(\varphi)}{3}\Big)\Tr(\bm{Q}^2) \\ 
- \frac{\eta(\varphi)}{3}\Tr(\bm{Q}^3) + \frac{\eta(\varphi)}{4}\Tr(\bm{Q}^2)^2\Big\} + \frac{1}{2}k_\varphi\left(\bm{\nabla}\varphi\right)^2 \\
+\frac{1}{2}K_1\left(\bm{\nabla} \cdot \bm{Q}\right)^2 + \frac{1}{2}K_C\left(\bm{\nabla} \times \bm{Q} + 2\left(\frac{2\pi}{\lambda}\right) \bm{Q}\right)^2 \\
- \epsilon_a\bm{E}\cdot \bm{Q} \cdot \bm{E} + L_0 \bm{\nabla}\varphi \cdot\bm{Q}\cdot \bm{\nabla}\varphi.
\end{split}
\label{eq:fe}
\end{equation}
The first term in Eq.~(\ref{eq:fe}) is the bulk free energy for a binary fluid, with minima at $\varphi = 0,1$, corresponding to the isotropic and cholesteric phases, respectively. The second term describes a first-order, isotropic-nematic transition at $\eta(\varphi) = 2.7$: for  $\eta(\varphi) < 2.7$  the nematic component of the binary fluid is in the  isotropic (disordered) state, while for $\eta(\varphi)>2.7$ it is in the nematic or cholesteric phase. The third term penalises gradients in the concentration and provides a surface tension that favours circular drops. Compared to previous work \cite{Metselaar2017} the elastic contribution to the free energy now includes a term dependent on $\lambda$ \cite{Wright1989}. The parameter $\lambda$ gives the pitch length and is related to the chirality of the cholesteric helix (we will consider $\lambda > 0$ corresponding to right-handed chirality). The limit $\lambda^{-1}=0$ and $K_1=K_C$ gives a nematic liquid crystal within the one elastic constant approximation. The penultimate term in Eq.~(\ref{eq:fe}) is the contribution from an external electric field $\bm{E}$, with $\epsilon_{a}$ the dielectric anisotropy, while the last term provides an interfacial anchoring. $L_0 > 0$ encourages the director to align parallel to the isotropic-nematic interface (planar anchoring), whereas $L_0 < 0$ encourages perpendicular (homeotropic) anchoring. The coefficients used in the simulations are listed in Table~\ref{tab:parameters}.
\begin{table}[tb]
\begin{center}
\begin{tabular}{|c|l|c|}
\hline 
\textbf{Symbol} & \textbf{Description} & \textbf{Value} \\ 
\hline 
$ K_1 $ & Bend/splay elastic coefficient & 0.06-0.08 \\ 
\hline 
$ K_C $ & Twist elastic coefficient & 0.04-0.16 \\ 
\hline 
$ \lambda $ & Characteristic pitch & 10-46 \\ 
\hline 
$ k_\varphi $ & Surface tension coefficient & 0.03-0.06 \\ 
\hline 
$ L_0 $ & Interfacial anchoring & 0.06 \\ 
\hline 
$ \epsilon_a $ & Dielectric anisotropy & 1 \\ 
\hline 
$ \xi $ & Alignment parameter & $ 0.7 $\\ 
\hline 
$ A_\varphi $ & Binary fluid coefficient &  $ 0.2 $ \\ 
\hline 
$ A_0 $ & Landau-de Gennes coefficient &  $ 1.5 $ \\ 
\hline 
$ \Gamma $ & Rotational diffusivity & $ 0.7 $ \\ 
\hline 
$ M $ & Mobility & $ 0.7 $ \\ 
\hline 
$ \rho $ & Density & $ 1 $ \\ 
\hline 
$ \nu $ & Viscosity & $ 2/3 $\\
\hline 
$ P_0 $ & Bulk pressure & $ 0.25 $
\\ \hline 
\end{tabular} 
\end{center}
\caption{Simulation parameters.}
\label{tab:parameters}
\end{table}

The evolution of the order parameters $\varphi$ and $\textbf{Q}$ is dictated by the Cahn-Hilliard and Beris-Edwards equations, respectively \cite{Cahn1958,Beris1994},
\begin{equation}\label{eq:CahnHilliard}
\partial_t \varphi + \bm{\nabla} \cdot \left(\varphi\bm{u}\right) = M \bm{\nabla}^2 \mu,
\end{equation}
\begin{equation}\label{eq:BerisEdwards}
\left(\partial_t + \bm{u} \cdot \bm{\nabla}\right)\bm{Q} - \bm{S} = \Gamma \bm{H}\ad{,}
\end{equation}
where $\bm{S} = \left(\xi \bm{D}+\bm{\Omega}\right)\left(\bm{Q}+\frac{1}{3}\bm{I}\right) + \left(\bm{Q}+\frac{1}{3}\bm{I}\right)\left(\xi \bm{D}-\bm{\Omega}\right)-2\xi\left(\bm{Q}+\frac{1}{3}\bm{I}\right)\Tr\left(\bm{Q}\bm{W}\right)$ characterises the response of the nematic tensor $\bm{Q}$ to velocity gradients \cite{Beris1994}. $\bm{D}$ and $\bm{\Omega}$ are the symmetric and antisymmetric parts of the velocity gradient tensor $\bm{W}=\bm{\nabla u}$, and $\xi$ is the tumbling parameter. $\Gamma$ is a rotational diffusivity and $\bm{H} = -\left(\frac{\delta\mathcal{F}}{\delta\bm{Q}} - \frac{1}{3}\bm{I}\Tr\frac{\delta\mathcal{F}}{\delta\bm{Q}}\right)$ is the molecular field, accounting for the relaxation towards a free energy minimum while $\bm{Q}$ remains traceless. In Eq.~(\ref{eq:CahnHilliard}) $M$ is the mobility and $\mu = \frac{\partial f}{\partial \varphi} - \bm{\nabla} \cdot\left(\frac{\partial f}{\partial (\bm{\nabla} \varphi)}\right)$ is the chemical potential.

The velocity $\bm{u}$ evolves according to the generalised incompressible Navier-Stokes equations
\begin{align}\label{eq:NavierStokes}
\nabla\cdot\bm{u}&=0,\\
\rho\left(\partial_t + \bm{u} \cdot \bm{\nabla}\right)\bm{u} &= \bm{\nabla} \cdot \bm{\Pi},
\end{align}
where $\bm{\Pi}$ is the stress tensor. The stress consists of a viscous stress $\bm{\Pi}_{\text{viscous}}=2\nu\bm{D}$, an elastic stress $\bm{\Pi}_{\text{elastic}}=-P_0\bm{I}-\xi\bm{H}\left(\bm{Q}+\frac{1}{3}\bm{I}\right)-\xi\left(\bm{Q}+\frac{1}{3}\bm{I}\right)\bm{H}+2\xi\left(\bm{Q}+\frac{1}{3}\bm{I}\right)\Tr\left(\bm{QH}\right)+\bm{Q}\bm{H}-\bm{H}\bm{Q}- \bm{\nabla Q}\delta f/\delta\bm{\nabla Q}$, and a capillary stress $\bm{\Pi}_{\text{capillary}}=\left(f-\mu\varphi\right)\bm{I}-\bm{\nabla}\varphi\left(\delta\mathcal{F}/\delta\bm{\nabla}\varphi\right)$, where $\nu$ is the fluid viscosity and $P_0$ is the bulk pressure.

Simulations are performed on a $L_x\times L_y\times L_z=100 \times100\times 1$ lattice with periodic boundary conditions in all directions. For bulk cholesteric systems the initial configuration is a single pitch rotating around the $y$-axis. For tactoids, the initial configuration for the simulation is a circular droplet with a uniform director field pointing in the $z$-direction.

\section{Phase diagrams}

\begin{figure}
  \includegraphics[width=\textwidth]{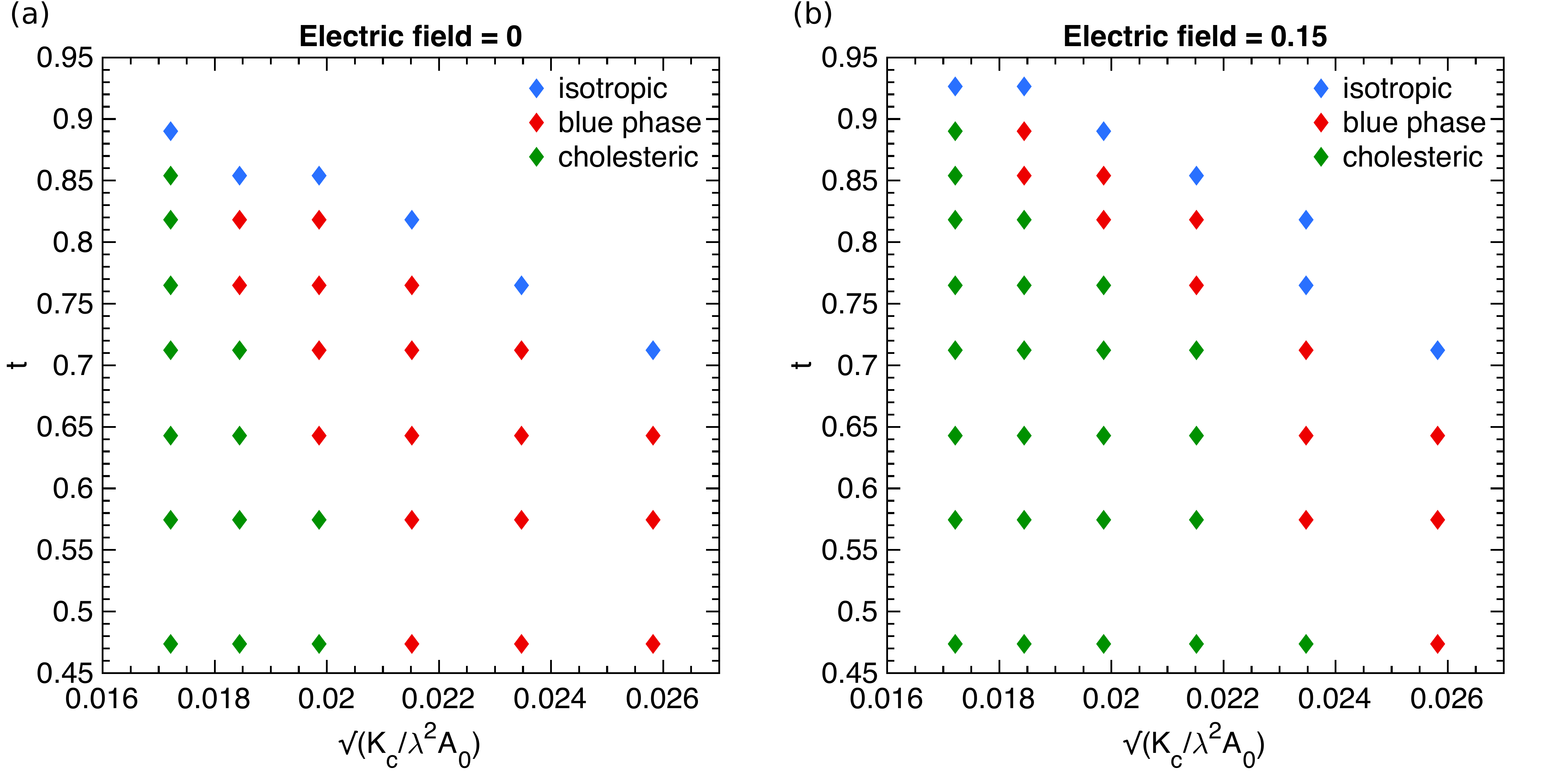}%
\caption{Phase diagrams for varying dimensionless pitch $\sqrt{K_C/\lambda^2A_0}$ and reduced temperature $t=27(1-\frac{1}{3}\eta)/\eta$. (a) Zero external field. (b) External field E = 0.15. In the presence of the field, the two-dimensional blue phase is suppressed in favour of the cholesteric phase.}
\label{fig:phase_diagrams}
\end{figure}

We begin by constructing phase-space plots, both without (Fig.~\ref{fig:phase_diagrams}a) and with (Fig.~\ref{fig:phase_diagrams}b) an external electric field, to illustrate the range of temperature and chiral strength ($\sqrt{K_C/\lambda^2A_0}$) for which isotropic, cholesteric, or blue phases form. For each combination of parameters we find the lowest free energy state by running the dynamical simulation. Fig.~\ref{fig:phase_diagrams}a shows a phase diagram in zero external field for a range of reduced temperatures and chiral strengths. The reduced temperature $t$ is defined as $27(1-\frac{1}{3}\eta)/\eta$ in order to have $t$ equal to $1$ at the isotropic-nematic transition. $K_1$ and $K_C$ are set to $0.08$ and $0.1$, respectively. 

As expected, at high temperatures the isotropic phase is stable, while at low temperatures the blue and cholesteric phases are formed. The cholesteric phase is favoured when the chirality is low (the pitch $\lambda$ is high). Upon increasing the chirality double twist becomes locally favourable leading to half-skyrmions. Topological defects form and organise in hexatic lattices to release the high global elastic energy penalty of the double twist, and a two-dimensional blue phase emerges.
 
Application of an external field along the $z$-axis shifts the ordering to that shown in Fig.~\ref{fig:phase_diagrams}b. As the electric field is increased, the preference of the director field to lie out of plane along $z$ suppresses double twist deformations and the half-skyrmion lattice becomes less stable, in favour of cholesteric patterns. Interestingly, at high temperatures and relatively small chiral strength the aligning field stabilises isolated half-skyrmions, separated by isotropic domains and without any $-1/2$ topological defects, which organise into a hexatic lattice. This behaviour is markedly different from known simulation results for hexagonal blue phases in fully three-dimensional systems \cite{Henrich2010} or confined between two parallel plates \cite{Fukuda2013}. For stronger electric fields the director field is out of plane in the entire domain and a (para)nematic phase, with alignment in the $z$-direction, is formed.

\begin{figure*}
  \includegraphics[width=1.0\textwidth]{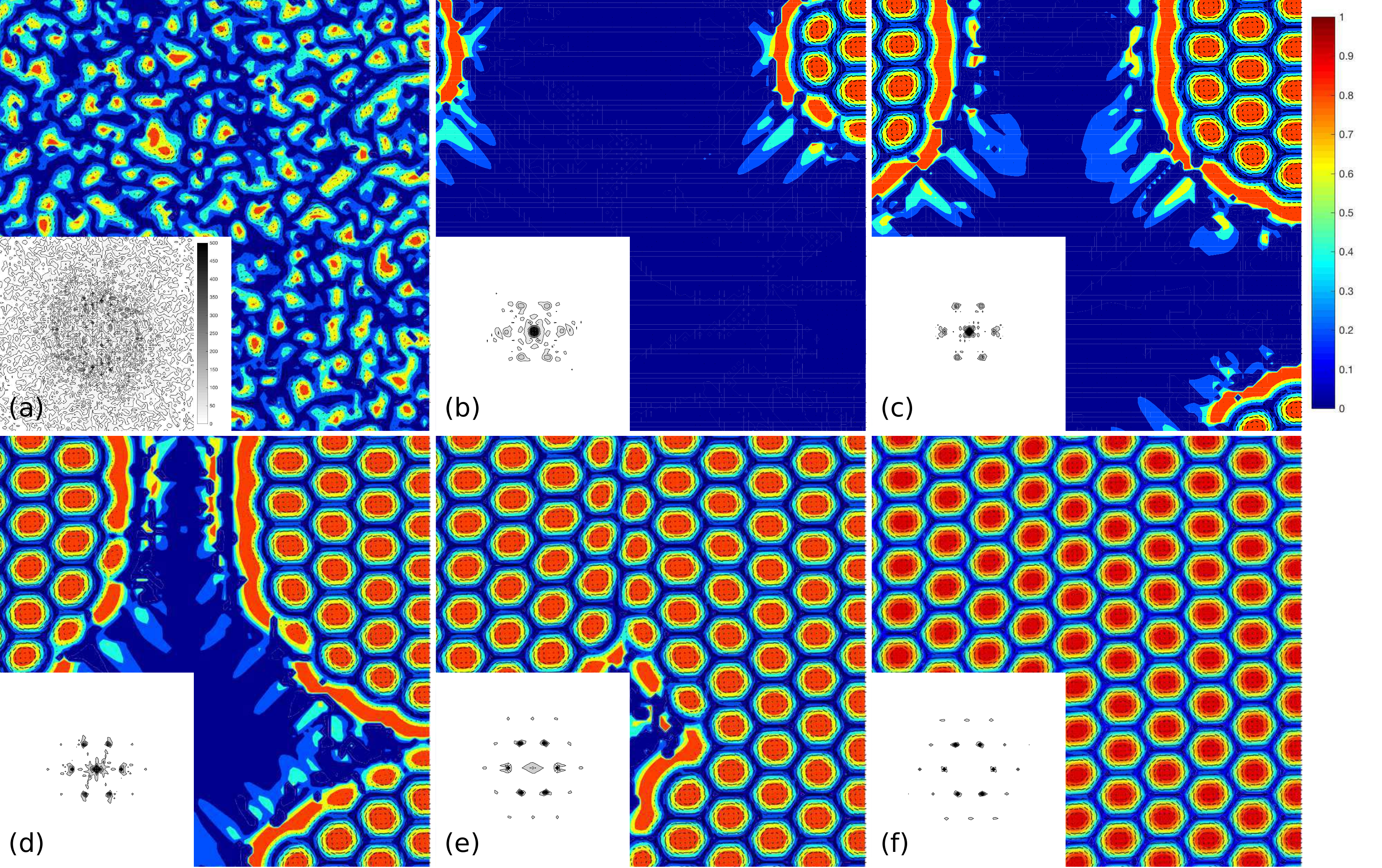}%
\caption{Real-space images of a quench from the isotropic phase into the two-dimensional blue phase. (a-f) are images taken 20,~400,~800,~1200,~1800 and 3000 simulation time steps after quenching. The colour coding indicates the absolute value of the projection of the director onto the direction perpendicular to the plane ($z$-axis). The insets in the lower left corners are Fourier transforms of the director field (see also Supplementary Movie I.)}
\label{fig:phase_transition}
\end{figure*}

In order to characterise the dynamics of the blue phase formation, we investigate a quench from the isotropic state into the two-dimensional blue phase. The simulation is set up in a random state, corresponding to infinite temperature, and the temperature is instantaneously lowered to $t=0.47$, while the chiral length $\sqrt{K_C/\lambda^2A_0}=0.0235$ is kept constant at a value where the blue phase is stable. 
Fig.~\ref{fig:phase_transition} shows snapshots tracking the temporal evolution of the transition to the blue phase, showing that it is formed through a nucleation and growth process (see also Supplementary Movie I). Measurement of the structure factor for the director field $S_{{\bf k}}=\sum_{{\bf k}}\text{exp}(i{\bf k}\cdot{\bf n})$, shows emerging hexatic order. Compared to previous theoretical and computational studies of the formation of 2D blue phases~\cite{Grebel83,Duzgun2017}, here we fully resolve hydrodynamic effects. Although the kinetic pathway shows minor differences, no qualitative changes in the dynamics of phase ordering are observed.

It is worth noting that the final hexagonal lattice is not perfect. This is because we do not include any thermal fluctuations in these simulations, so the system cannot overcome energy barriers, and due to the finite size of the simulation box, which leads to a small mismatch with the preferred period of the blue phase.

\section{Chiral nematics in confinement}

So far we have considered only bulk cholesteric and blue phases. We now focus on the impact of confining interfaces, and consider the morphology of liquid crystal tactoids by simulating droplets of chiral liquid crystal within an otherwise isotropic fluid medium. In order for the droplets to feel the confinement, the interfacial anchoring has to be non-zero, as having no anchoring would be equivalent to slicing a circular region out of an infinite domain. Therefore, in the simulations a choice has to be made for either planar or homeotropic interfacial anchoring. Here, as predicted in theory and simulations of isotropic-nematic interfaces for rod-like systems \cite{VanDerSchoot1999,McDonald2000} and in the experimental realisations~\cite{Chen1998}, the anchoring is chosen to be strongly planar at the interface.

\begin{figure}
  \includegraphics[width=0.9\textwidth]{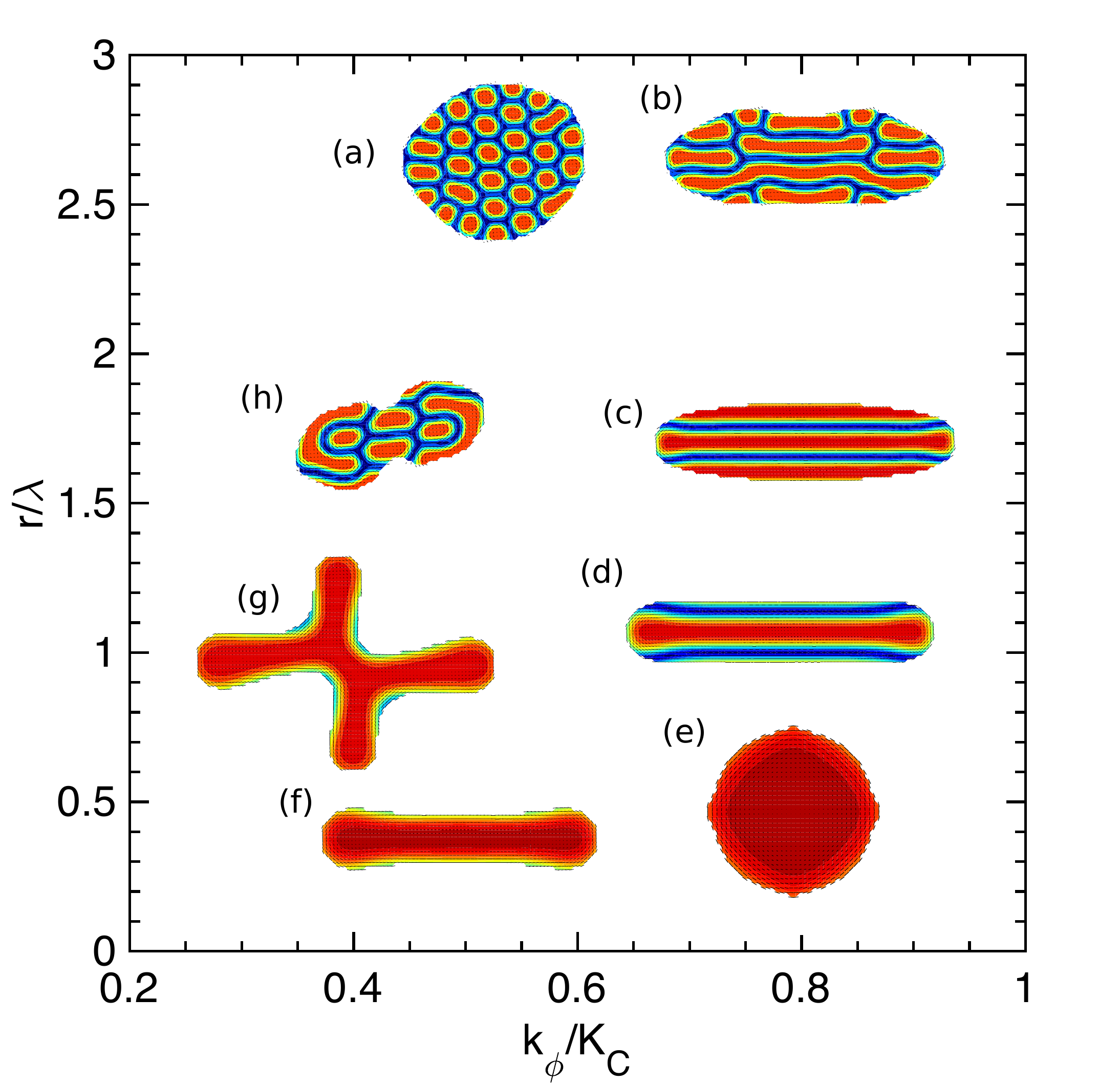}%
\caption{Phase diagram for cholesteric tactoids. The initial condition in all cases is a circular droplet with the director pointing in the $z$-direction and the bend-splay elastic coefficient is equal to $0.06$ in simulation units. The colour coding indicates the absolute value of the projection of the director onto the direction perpendicular to the plane ($z$-axis), and black lines are the projections of the director onto the plane. The variables are droplet radius, $r$, cholesteric pitch, $\lambda$, surface tension coefficient, $\kappa_{\phi}$, and twist elastic coefficient, $K_C$.
(a) $r = 24$, $\lambda = 10$, $K_C = 0.12$, $k_\phi = 0.06$: two-dimensional blue phase, but with defects in the lattice structure due to confinement.
(b) $r = 24$, $\lambda = 10$, $K_C = 0.08$, $k_\phi = 0.06$: an elongated droplet with single-twisted domains.
(c) $r = 18$, $\lambda = 15$, $K_C = 0.08$, $k_\phi = 0.06$: an elongated tactoid with cholesteric stripes. 
(d) $r = 24$, $\lambda = 24$, $K_C = 0.08$, $k_\phi = 0.06$: a single cholesteric stripe in an elongated tactoid.
(e) $r = 18$, $\lambda = 36$, $K_C = 0.04$, $k_\phi = 0.03$: ordering predominantly along $z$.
(f) $r = 18$, $\lambda = 36$, $K_C = 0.16$, $k_\phi = 0.06$: a single cholesteric stripe.
(g) $r = 24$, $\lambda = 36$, $K_C = 0.16$, $k_\phi = 0.06$:  a metastable star shape.
(h) $r = 18$, $\lambda = 15$, $K_C = 0.16$, $k_\phi = 0.06$: strong deformation and local double twist cylinders.}
\label{fig:membranes_diagram}
\end{figure}

Choosing as axes the ratio of the droplet radius $r$ to the cholesteric pitch $\lambda$ and the ratio of $k_{\phi}$, which controls the surface tension, to $K_C$, the twist elastic coefficient, Fig.~\ref{fig:membranes_diagram} indicates the structures found in different regions of parameter space. The initial condition in all cases is a circular droplet with the director pointing in $z$-direction: this could be achieved in experiment by applying a strong out-of-plane external field.

Panel (a) shows that for large $r/\lambda$ and large twist elastic coefficient the system relaxes into a two-dimensional blue phase. Irregularities in the lattice structure near the boundaries ensure that the tactoid satisfies the planar boundary conditions without the elastic energy penalty that would be associated with strong local curvature of the interface. In panel (b) the twist elastic coefficient is smaller, which means that uniaxial cholesteric ordering becomes more favourable relative to the double twist of the blue phase. The smaller twist elastic coefficient compared to the bend-splay elastic coefficient also encourages tactoid elongation, so that the system forms a uniaxial structure in the centre, but with distorted half-skyrmions near the boundaries to better satisfy the anchoring condition. As $r/\lambda$ is decreased, panels (c) and (d), uniaxial chiral ordering dominates and the droplet is strongly stretched along the chiral axis to satisfy the boundary conditions with minimal elastic energy penalty.

In panel (e), when the pitch becomes even longer compared to the droplet size, and the twist elastic coefficient is small, there is a change in behaviour, and it becomes favourable for the director to point out of plane, with a small twist at the surface. There is therefore no reason for the tactoid to elongate.

For larger twist elastic coefficient and smaller surface tension, the twist free energy dominates and the system forms a single cholesteric stripe, elongating to give the correct pitch over the short axis of the tactoid (panel (f)). For a slightly larger droplet, the stable state is a longer stripe. However, a metastable star-shape (panel (g)) can also form from a configuration where the nematogens are initially out of plane. This results from the initial dynamics which is a two dimensional twist. It is worth noting that such a metastable configuration is not possible without including hydrodynamic effects. Finally, in panel (h), double twist regions become more favourable again, and the low surface tension allows deformations so that these can adopt the correct pitch, subject to satisfying the boundary conditions.

\begin{figure*}
  \includegraphics[width=\textwidth]{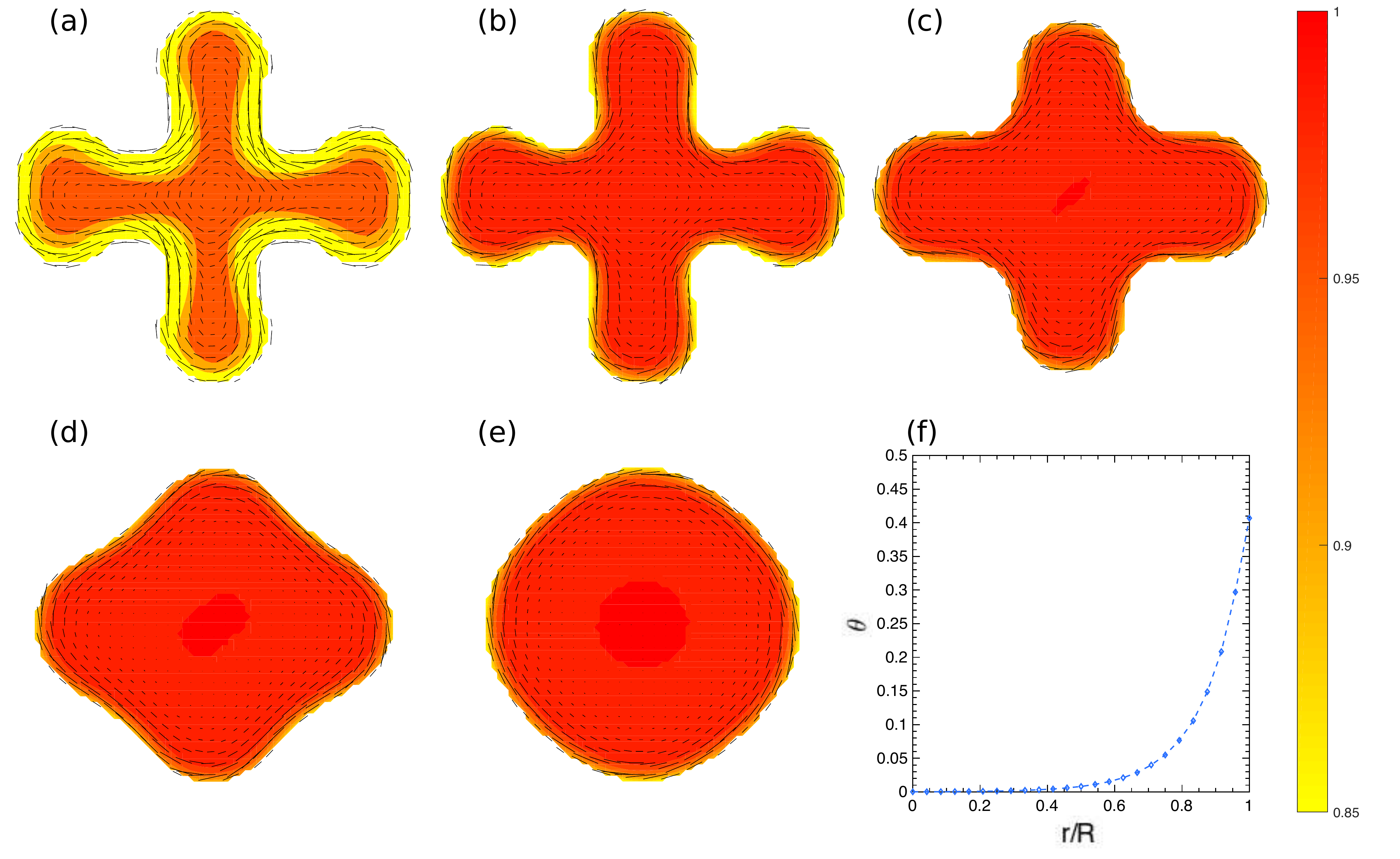}%
\caption{Time evolution of a star shaped tactoid after applying an external field. (a-e) Snapshots after $0,~200,~500,~2000$ and $12000$ simulation time steps.  The colour indicates the value of the $z$-component of the director. (f) Angle of the director with respect to the $z$-axis 12000 time steps after application of the field, as a function of the distance from the centre of the tactoid to the edge, showing increasing twist approaching the edge.}
\label{fig:dogic}
\end{figure*}
\section{Chiral tactoids in an external field}

The tactoids in Fig.~\ref{fig:membranes_diagram} are all stable (or metastable) in the absence of an external field. By applying a sufficiently strong field in the $z$-direction it is possible to align the directors out of plane. In Fig.~\ref{fig:dogic} we follow the time evolution of a star-shaped tactoid after applying an external field. As the director aligns with the external field surface tension pulls the tactoid into a circular shape. Note, however, that there is still twist around its edges (Fig.~\ref{fig:dogic}e-f). This occurs because at the middle of the tactoid there is a competition between the aligning field and the Frank elasticity on the one hand, and the preferred chiral twist on the other. For a sufficiently strong field this leads to directors pointing along $z$. However, at the edge of the tactoid the relative contribution of the Frank elastic energy is less, leading to the possibility of twist.

Such an edge structure draws analogy to experimental observations of `chiral rafts' in dilute suspensions of rod-like, fd-viruses with attractive interactions induced by the addition of non-adsorbing polymers~\cite{Gibaud2012}. In these experiments the role of the non-adsorbing polymers is to provide attractive forces through the depletion mechanism between the otherwise repulsive rods. As a result of the depletion it is energetically favourable for the viruses to stand vertically except around the edge of the raft where twist deformations reduce the interfacial energy by lowering the polymer-virus interaction area. In our simulations, an electric field in the vertical direction mimics the depletion interaction, while at the free edges twisting results from the balance of chiral and elastic terms.

\section{Summary and outlook}

We have used nematohydrodynamic simulations of liquid crystals to study the order in, and the dynamics of, 2D blue phases. In particular, we have shown that 
a wealth of stable morphologies can be achieved by cholesteric tactoids, liquid crystal droplets within an otherwise isotropic liquid. Bulk 2D blue phases have been observed experimentally, and isolated skyrmions and hopfions have been created in cholesteric liquid crystals using optical excitations~\cite{Smalyukh2010,Ackerman2015}. Recently, confinement of cholesteric liquid crystals within 3D microchannels~\cite{Guo2016,Nych2017} and 3D droplets~\cite{Posnjak2016} have proven successful in observing both half- and full-skyrmions without application of external stimuli. Introducing further lateral confinement in such setups could reproduce various of the stable structures reported in this study.

Another potentially relevant experimental system is the rafts of fd-virus discussed above~\cite{Gibaud2017}. However, due to the strong depletion interactions in these experiments it has not been possible for the rods to lie in the plane of the surface, and thus cholesteric striped, stars or blue phase rafts have not been observed so far. Carefully tuning the strength of the depletion interaction may make it possible to access other parts of the phase space considered here. Tuning of the pitch could then be achieved through manipulation of genetic or physical cues in the experiments~\cite{Gibaud2017}. It would also be interesting to test for the effects of external stimuli such as an electric or magnetic field on the morphology of chiral rafts. From the theoretical point of view, one could envisage that applying the external aligning field at an angle with respect to the vertical direction could break the symmetry and lead to the formation of further exotic and controllable conformations. Similarly, applying mechanical forces could be a potential mechanism for building chiral macrostructures as was partly explored for chiral rafts, where applying extensional forces by optical traps was shown to create double or triple helices in the experiments~\cite{Gibaud2017}.

Finally, the next step in studying exotic tactoids is to investigate the interaction between different morphologies and the coalescence of such tactoids to construct hierarchical structures. Such structures have already been shown to successfully form from coalescence of chiral rafts of fd-viruses~\cite{Sharma2014}.

\section{Acknowledgements}
This project has received funding from the European Union's Horizon 2020 research and innovation programme under the DiStruc Marie Sk\l{}odowska-Curie grant agreement No. 641839. AD was supported by a Royal Commission for the Exhibition of 1851 Research Fellowship. We thank Maxime Tortora for helpful conversations and help with software optimisation.
 
\bibliographystyle{apsrev4-1}
\bibliography{chirality}

\end{document}